\documentclass[a4paper]{VisionStyle}
\usepackage{epsfig}

\begin{document}

\title{In-orbit calibration of vignetting in EPIC}

\author{D.H.\,Lumb\inst{1} } 

\institute{
  ESA Payload Technology Division, Research and Scientific Support Department, ESTEC,   Postbus 299, NL-2200 AG Noordwijk, The Netherlands}

\maketitle 

\begin{abstract}
We briefly describe the in orbit measurements of the mirror vignetting, using observations
of SNR G21.5-09. The instrument features which complicate these measurements are briefly
described, and we show the spatial and energy dependences, outlining assumptions made
in deriving the eventual agreement between theory and measurement

\end{abstract}

\section{Introduction}\label{dlumb-WA2c_sec:intro}
The reduction in effective area with radial distance from the field of view centre,
or vignetting, must be accurately determined to support a number of important science
investigations:\begin{itemize}
\item Extended targets (e.g. clusters) whose radial brightness distribution must be accurately
traced in order to determine mass
\item Population studies - where exposure maps and counts to flux conversions depend upon
vignetting correction
\item Background studies of diffuse cosmic radiation where the normalisation for 
integrated flux over large areas of sky must be determined 
\end{itemize}

Direct measurement on-ground was prevented because all X-ray beam measurements were
performed in a non-parallel beam. The installation of an X-ray stray-light baffle and the
RGA stack introduced potential complications that could only be followed in the visible
light at the EUV parallel beam facility, so that X-ray energy dependence was not measureable.
Although the measured geometric vignetting factor was comparable with predictions, it was necessary to 
use in-orbit data to confirm the energy dependence, and further check that the geometric
factor was maintained through the spacecraft AIV and launch campaigns.

In essence, to measure the vignetting we need to measure a compact, simple spectrum
non-variable source at locations off-axis, and compare the inferred spectrum with that
of the same object measured on-axis. Truly point sources with reasonable brightness are
precluded because the effects of pile-up are severe, and furthermore vary with the off-axis
PSF changes, as well as the count rate reduction due to the vignetting itself. Extended objects require a complex
ray-tracing and PSF-folding to account properly for the vignetting component. While a
number of viable targets were selected for the in-orbit calibration, we have concentrated
on G21.5-09 for this work (see also \cite{dlumb-WA2c:bob}):
\begin{itemize}
\item It is moderately compact (core slightly larger than the PSF FWHM but $\leq$1 arcmin)
\item Absorbed simple power law spectrum
\item Count rate just below the on-axis pile-up limit
\end{itemize}
\section{Observation Set up}\label{dlumb-WA2c_sec:obsv}
The initial choice of pointing locations was complicated by the need to ensure that no 
significant portion of the remnant fell near CCD gaps. Given the orthogonal orientation
of the two MOS cameras, together with the totally different gap patterns in the PN, this
severely constrained the orientation avaialable, and  an angle $\sim$7 degrees off the nominal chip axes, and a
field angle of 10 arcminutes was chosen. 

As a consequence of the grating array angles and blocking fraction, the vignetting in the MOS
cameras is expected to be a strong function of azimuthal angle, so 4 locations were
scheduled to sample the extreme ranges of RGA blocking. A detailed simulation of expected
source parameters indicated that $\sim$30ks exposure per location was needed to measure
the vignetting with adequate leverage to highest energies.

\begin{figure}[ht]
  \begin{center}
    \epsfig{file=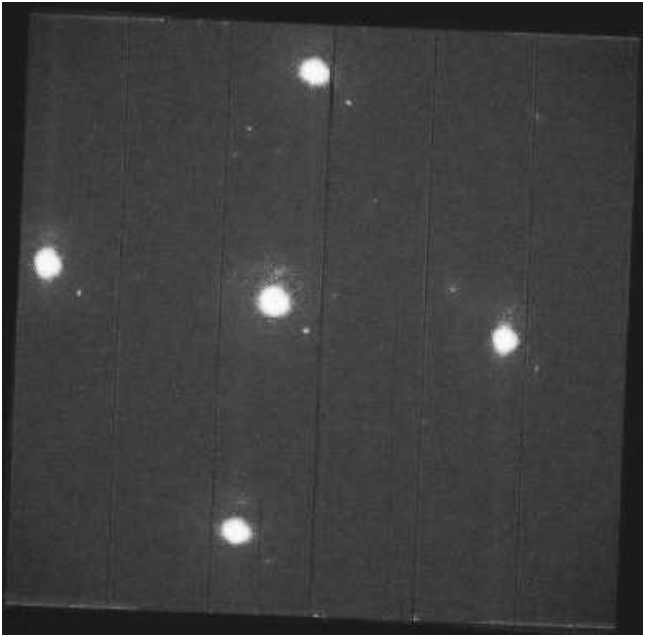, width=8.5cm}
  \end{center}
\caption{Merged image of the 5 major pointings made on G21.5-09}  
\label{fauthor-E1_fig:G21image}
\end{figure}
\section{Initial Analysis}\label{dlumb-WA2c_sec:anal}
The observations were conducted in April 2000 during a period of extended proton flares.
Not only did this curtail usable exposure duration to $\sim$5 -- 20 ks per exposure, but even
the quietest periods suffered higher than usual quiescent background. Therefore
considerable effort was expended on understanding the effects of background subtraction 
using in-field areas close to the target. 

Our understanding of the MOS azimuthal variations in vignetting were undermined by significant
variations ($\sim$10\%) in relative vignetting measured in the PN camera, from azimuth to
azimuth. This was attributed to a combination of incomplete background correction and to discrepancies
in the exposure calculations influenced by the higher than nominal background. 

Eventually it was realised that these relative variations were correlated with camera
orientations, and traced back to similar unresolved discrepancies between mirror alignment 
lens and inferred telescope axes measured at the EUV test facility (\cite{dlumb-WA2c:CSL}). At the time these oientation discrepancies were claimed to be
irreproducible to $\sim$ 20 arcsec level, but were also seen in similar magnitude and
direction in the Panter calibration of maximum throughput orientation (\cite{dlumb-WA2c:PAN}). 

For the in-orbit data,
acceptable agreement between predictions and inferred vignetting value could be obtained 
by positing an offset between the nominal
telescope axis intersection at the focal plane, and the actual location of {\em both} the nomainal
telescope axis and the boresight axis of the XMM system which determines the location of 
the central target. 

Under the assumption of azimuthal symmetry (although this is not necessarily valid
 due to the mechanical tolerances on mounting the X-ray baffle), we compare the counts
per energy bin between {\em pseudo} on- and off-axis locations as measured on G21.5-09, and
the relative vignetting between the corresponding locations. The energy bins' widths were
varied semi-logarithmically to maintain reasonable signal:noise per bin.

\section{Results}\label{dlumb-WA2c_sec:results}
\begin{figure}[ht]
  \begin{center}
    \epsfig{file=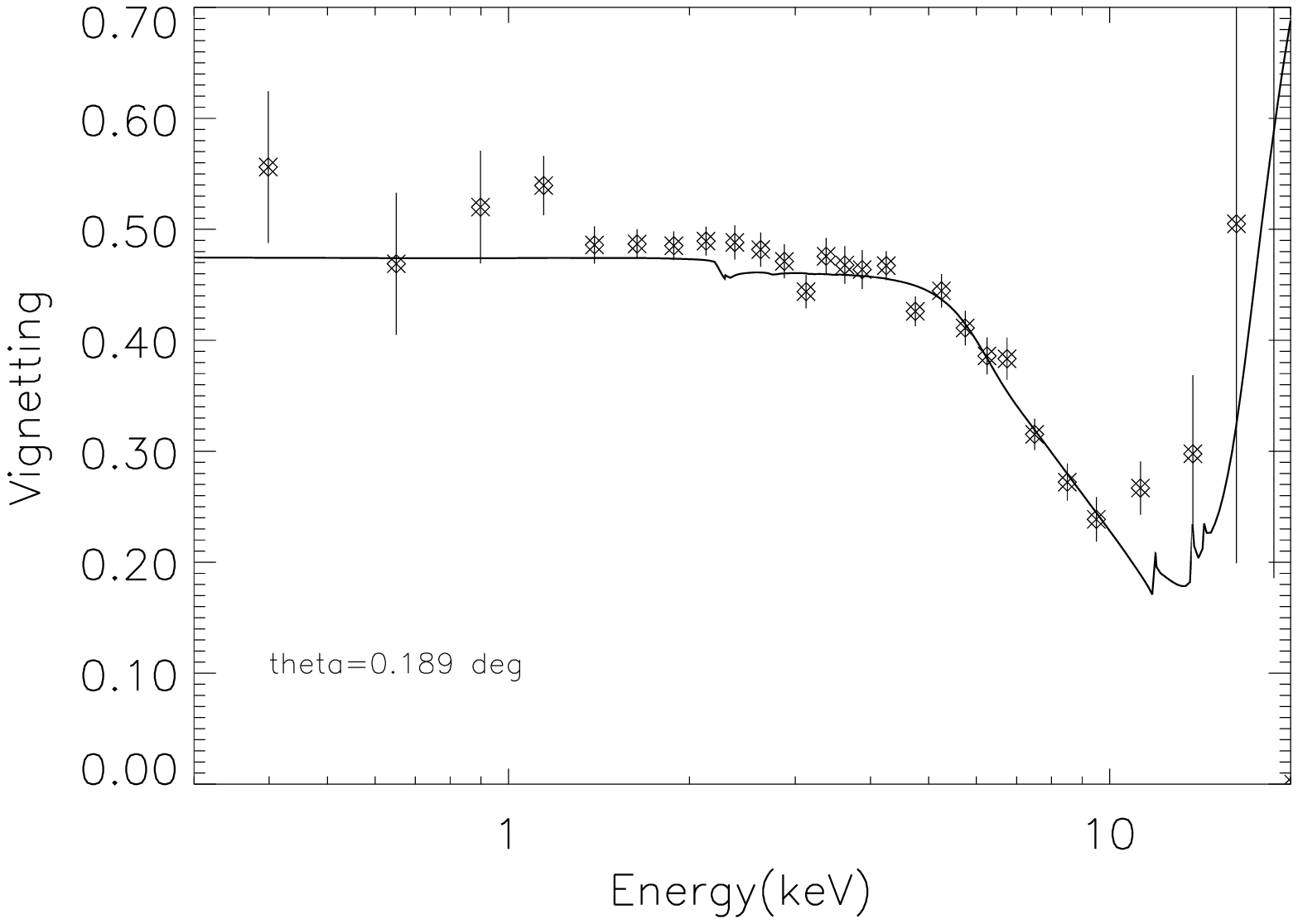, width=9.2cm}
  \end{center}
\caption{{\em Relative} vignetting of the PN telescope for an off-axis angle of11.3 arcminutes, compared with the nominal boresight location}  
\label{fauthor-E1_fig:PNR65}
\end{figure}

A single azimuth vignetting measurement for the PN camera, in the lowest background
exposure, is shown in Figure~\ref{fauthor-E1_fig:PNR65}. The energy at which
the vignetting decreases strongly is determined by the critical angle for grazing incidence
at the off-axis angle of the target (11 arcminutes in this case). The increase again at higher energies is a consequence of only the innermost mirror shells providing substantial
reflectivity. For a small diameter shell, at high energies, the area
increases intially with off-axis angle: on one side
of the mirror the parabola grazing angle is shallower than for the on-axis geometry.
The corresponding hyperbola graze angle is then larger but because of the
asymmetry of the reflectance vs. angle curve the higher reflectivity on the
parabola dominates the product of the reflectances.   
A subsequent calibration set in  ca. April 2001 at larger off-axis angle
allowed some measure of sensitivity to the change in $\theta$, and confirms
the validity of the model.
\begin{figure}[ht]
  \begin{center}
    \epsfig{file=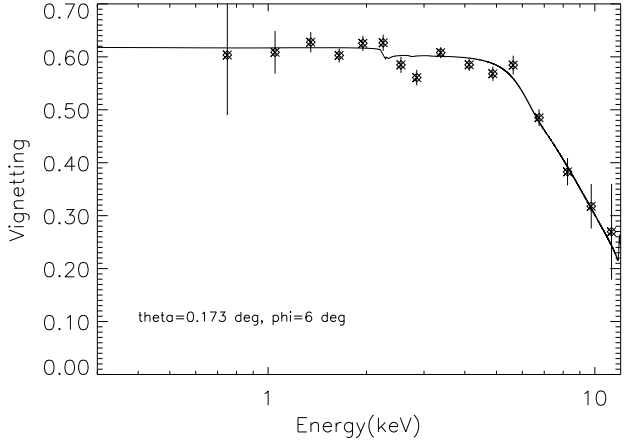, width=9.2cm}
  \end{center}
\caption{{\em Relative} vignetting of the MOS1 telescope for an off-axis angle of 10.4 arcminutes, compared with the nominal boresight location}  
\label{dlumb-WA2c_fig:M1R62}
\end{figure}

A comparable vignetting measurement for the MOS cameras is shown in Figure~\ref{dlumb-WA2c_fig:M1R62}. Due to the lower
effective area of the MOS cameras, the S:N is lower than the PN camera, and the energy bins are wider. It was found again that there
was a potential telescope axis misalignment. However records of the tests in ground facility
were not so clear, because the installation of the RGA had blocked the access to the mirror
alignment lens for most tests. Relying purely on inferred alignment of the axis based on the
vignetting itself undermines the goal of directly measuring the effect of RGA azimuthal
blocking factor.
\section{Energy Dependence}
\begin{figure}[ht]
  \begin{center}
    \epsfig{file=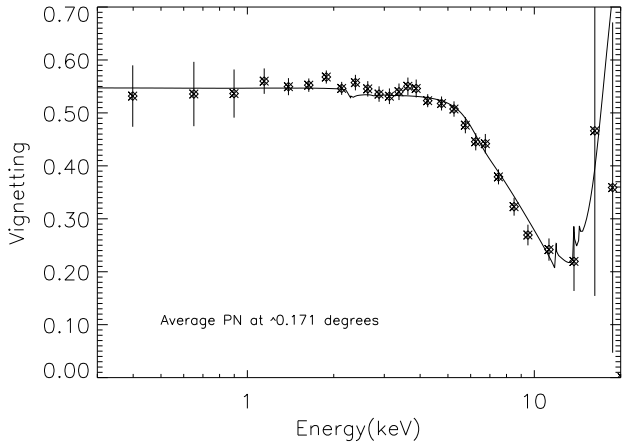, width=9.2cm}
  \end{center}
\caption{{\em Relative} vignetting of the PN telescope after averaging all azimuths around 10.3 arcmins off-axis. The energy dependence is in good agreement}  
\label{dlumb-WA2c_fig:PNAV}
\end{figure}
The PN data are relatively close in off-axis angle, and have no intrinsic azimuthal
dependence, so we should be able to average the 4 separate locations to check the 
predicted energy dependence is correctly reproduced. This is shown in Fig.~\ref{dlumb-WA2c_fig:PNAV}.

For the MOS data repeating the exercise is not really valid, given the large
variation in RGA blocking with azimuth. However to discern if the placement
of RGA gratings and ribs upsets the ``grey'' filter properties via. differential
shadowing of some sub-sets of shells, we nevertheless form the same average 
response in the 2 MOS cases. There seems to be no significant energy dependent
discrepancies. 
\begin{figure}[ht]
  \begin{center}
    \epsfig{file=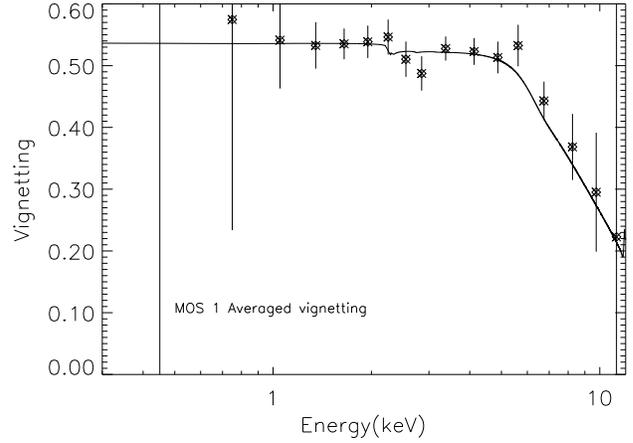, width=9.2cm}
 
    \epsfig{file=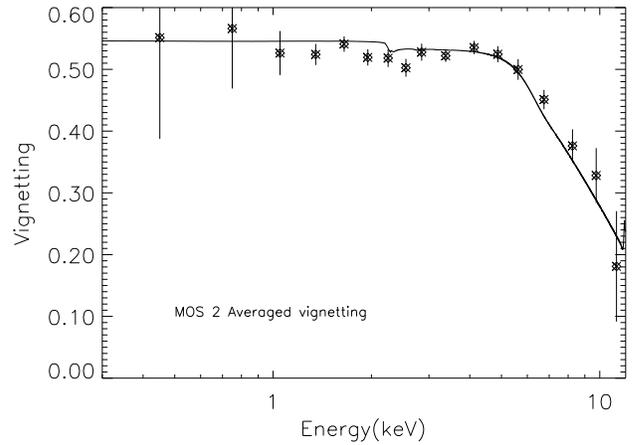, width=9.2cm}
  \end{center}
\caption{{\em Relative} vignetting of the MOS telescopes after averaging all azimuths around 10.3 arcmins off-axis. The energy dependence is in good agreement}  
\label{dlumb-WA2c_fig:MOS2AV}
\end{figure}
\section{Conclusions}
The energy dependent vignetting calibration can be well matched to
pre-launch predictions, but only on an assumption that the telescope optical
axis is not well-alligned with the telescope boresight. This is not unexpected
following difficulties on-ground of maintaining and/or measuring the telescope
axis to better than 10's arcseonds.
 
We note finally that the assumed telescope axis misalignment implies that ``on-axis'' 
targets   at the common boresight location are actually at a slightly different vignetting
value per telescope. We speculate that this partly accounts for some of the observed flux 
discrepancies between the MOS and PN cameras.

\section{Acknowledgements}
Fruitful discussions with C Erd, P Gondoin, A Finoguenov and D Neumann are
gratefully acknowledged


\begin{thebibliography}{}
\bibitem[\protect\astroncite{Egger et al}{1997}]{dlumb-WA2c:PAN}
Egger, R., Aschenbach, B., Br\"auninger et al 1997 Panter Test Reports XMM-TS-PA063/970829, XMM-TS-PA070/980115 etc.

\bibitem[\protect\astroncite{Stockman et al}{1997}]{dlumb-WA2c:CSL}
Stockman, Y., Tock, J-P., Thome, M. et al 1997 CSL Test Reports, RP-CSL-MEV-97032, RP-CSL-MEV-97019 etc.

\bibitem[\protect\astroncite{Warwick et al}{2001}]{dlumb-WA2c:bob}
Warwick, R.S., Bernard, J-P., Bocchino, F. et al 1996, A\&A 365 L248.


\end{thebibliography}
\end{document}